\documentclass[runningheads]{llncs}
\usepackage{graphicx}
\usepackage{caption} 
\begin{document}
\title{A Survey on Common Threats in npm and PyPi Registries}
%
%\titlerunning{Abbreviated paper title}
% If the paper title is too long for the running head, you can set
% an abbreviated paper title here
%
\author{Berkay Kaplan\orcidID{0000-0002-4365-7606} \and Jingyu Qian\orcidID{0000-0002-3953-5382}}

\institute{University of Illinois at Urbana-Champaign, Urbana, IL 61801-2302, USA
\email{\{berkayk2, jingyuq2\}@illinois.edu}
}

\maketitle              % typeset the header of the contribution
\begin{abstract}
Software engineers regularly use JavaScript and Python for both front-end and back-end automation tasks. On top of JavaScript and Python, there are several frameworks to facilitate automation tasks further. Some of these frameworks are Node Manager Package (npm) and Python Package Index (PyPi), which are open source (OS) package libraries. The public registries npm and PyPi use to host packages allow any user with a verified email to publish code. The lack of a comprehensive scanning tool when publishing to the registry creates security concerns. Users can report malicious code on the registry; however, attackers can still cause damage until they remove their tool from the platform. Furthermore, several packages depend on each other, making them more vulnerable to a bad package in the dependency tree. The heavy code reuse creates security artifacts developers have to consider, such as the package reach. This project will illustrate a high-level overview of common risks associated with OS registries and the package dependency structure. There are several attack types, such as typosquatting and combosquatting, in the OS package registries. Outdated packages pose a security risk, and we will examine the extent of technical lag present in the npm environment. In this paper, our main contribution consists of a survey of common threats in OS registries. Afterward, we will offer countermeasures to mitigate the risks presented. These remedies will heavily focus on the applications of Machine Learning (ML) to detect suspicious activities. To the best of our knowledge, the ML-focused countermeasures are the first proposed possible solutions to the security problems listed. In addition, this project is the first survey of threats in npm and PyPi, although several studies focus on a subset of threats.

\end{abstract}
\section{Introduction}
%% TODO: let's think about how to include ML in introduction.

The OS movement allows any developer to read or attempt to contribute to the source code hosted in a publicly accessible location, such as GitHub. This trend may have several advantages, such as code reviewers catching software bugs promptly. Furthermore, the creativity behind a software’s design is no longer limited to a team of developers as anyone is free to propose their ideas. This advantage allows additional features or functionalities to be implemented, thus possibly making the project the market leader in its area. 

Although several advantages are present with the OS movement, there are also disadvantages. One possible disadvantage is that certain individuals in public may have malicious intentions, and the publicly accessible code may raise security concerns. A team of founders of the OS platform may review every pull request to ensure the benign intentions of the commits. However, some environments are known to lack control measures. Some of these environments may include public registries, which are public repositories anyone can upload their packages to for others to use. These repositories intend to serve as collections of libraries to facilitate development tasks with readily available packages. Some public registries do not have any control measurements besides reporting the malicious code once detected by a developer. This issue gives the attacker a certain lifespan to perform malicious activities. 

PyPi and npm are two popular public registries. These environments require developers to verify their email addresses without further control, making it much easier for attackers to create malicious accounts. These accounts can publish any package in their public registry, amplifying an attack's impact naturally. Furthermore, attackers can take advantage of several other techniques that benefit from human errors. Some of these techniques are typosquatting and combosquatting. When an attacker intentionally publishes a package in the public registry with a similar name to popular packages, these techniques benefit from a potential typo during the manual installation of dependencies \cite{vu2020typosquatting}. Typosquatting takes advantage of typos during the installation of packages, while combosquatting hopes to exploit a mistake of the order in the package name, where the name consists of multiple nouns, such as "python-nmap" typed as "nmap-python" \cite{vu2020typosquatting}. The ultimate goal in both attacks is to get the developer to install the malicious package.

Another popular attack surface for these registries is the interdependent nature of its packaging environment. A handful of accounts maintain numerous packages. These packages are dependent on several other libraries. Compromising these accounts would give an attacker the ability to modify the code at will. Furthermore, if there is a vulnerability in one of the popular packages, then the issue will propagate through all the other libraries that depend on it. This concept is also known as software supply-chain security, which refers to outsiders accessing the final project \cite{ellison2010evaluating,ohm2020backstabber}. Software supply-chain security is another research field but seems to be heavily related to OS registry security. There has already been a study on software supply-chain and typosquatting attacks on public registries that includes npm and PyPi; however, the study did not cover other common dangers in the OS registries such as trivial packages \cite{duan2021towards}.

Another attack surface in the OS settings is the technical lag, which is the time or version difference between the dependency used in production and the most recent version available for installation \cite{zerouali2019impact}. For instance, when package A depends on several other libraries in production that are not up to their latest versions, known vulnerabilities may compromise package A as well through these dependencies. npm does not seem to help developers keep up with the latest versions of their libraries, as package.json requires developers to use a specific version or range of versions of a package instead of automatically choosing the most recent version \cite{zimmermann2019small}. 

We are the first to present a systematic survey of public registries' common vulnerabilities. In this paper, we will be exploring the impact of each vulnerability. We will be suggesting untested risk mitigation strategies we compiled from the literature to the administrators of the mentioned OS registries. We specifically incorporated ML techniques into our risk mitigation strategies to make our solutions more adaptable to different types of future attacks. We made this assumption as the features of ML-based tools can be changed to modify the program's purpose.

To the best of our knowledge, our work is the first in the field to offer countermeasures against the security problems in npm and PyPi environments. We have not conducted physical experiments or surveyed developers, as we merely focused on presenting the state of the field. We outlined our main contributions to the field:
%% TODO: might later add more contributions
\begin{itemize}
\item Compilation of all the threats and vulnerabilities from the literature that exists in public registries.
\item Compilation of work that explains the prevalence of the presented risks in popular OS environments.
\item Countermeasure suggestions against the mentioned attack surfaces.
\end{itemize}
Similar work has been published previously, such as the study from Vaidya et al. in 2019. The paper indeed built a systematic threat model in OS environments and provided threat classifications. Nevertheless, we determined that this study did not provide sufficient coverage as certain issues, such as trivial packages, were not discussed.

The OS movement has brought several advantages, and we would like to support the safety of its members. Supporting the public registry safety will also benefit the OS community as it would draw some companies, which could not enter the environment due to their high-security requirements. This paper aims to disclose the security issues in public registries to contribute to the OS community. This project's scope does not extend to vulnerabilities in native code or compiler as we focus more on software reuse.

\section{Background}

It would be beneficial to explain a couple of frequently used terms in this paper that relate to the interdependent structure of libraries.

\subsection{Dependencies and Dependency Trees}
%% TODO: Should dependencies be more like relationship instead of packages? I think the definition here is not precise.
Dependencies, packages, or libraries are fairly similar terms in the context of our survey. Dependencies are readily available code for developers to use in their projects to accelerate the speed of the implementation phase. Private companies may create these dependencies, or they may even be OS. The implementation in these libraries is usually not individually scrutinized for efficiency purposes in a project's life cycle.

A dependency tree visualizes the direct relationship between libraries. The simplest tree structure would be that package A depends on package B, meaning that package A uses components, functions, or code snippets from package B. These relationships can extend to several packages vertically, introducing more attack surfaces for a specific dependency. 

\subsection{Package Manager}
Package managers allow developers to manage their dependencies in a central document, for example, a JSON file. The managers facilitate the installation and sharing of dependencies as the developer no longer has to visit the library's official website to download the package into his local repository \cite{duan2020measuring}. The central document has a list of the dependencies. It is already connected to a registry that contains all the libraries needed via software. These managers are also configurable to connect to private registries, where all the libraries are internal to a company. Recent studies have shown that package managers have been abused to distribute malware, thus making them a trade-off between convenience and security \cite{duan2020measuring}. \cite{vaidya2019security}. 

\subsection{Software Supply-Chain Attacks}

The goal of supply-chain attacks is to inject malicious code into software or library externally. These attacks modify the targeted program so that it is still validly signed by its owner. The attacker simply injects his code into one of the software's dependencies \cite{ohm2020backstabber}. Theoretically, an attacker can do this injection in any given node in the software's dependency tree. The ultimate goal is to alter the behavior of the root or a specific node, which is the targeted software product, using a child node, which in this case is a dependency node in the tree \cite{ohm2020backstabber}.

\subsection{Typosquatting and Combosquatting}

Typosquatting is an attack methodology where an attacker intentionally publishes a package with a typo in its name, making the name relatively close to a popular package's name. The attacker hopes that a victim would make a typo while downloading dependencies. Thus, the victim would download the malicious package instead of the intended one. One example of this type of attack is the popular \textit{lodash} package incident, where it had a typosquatting package named \textit{loadsh}. In the English language, a developer might likely download the \textit{loadsh} package after making a typo.

Combosquatting is similar to typosquatting; however, some packages consist of multiple words in their name. Sometimes, it would be easy to forget the order of the words, and the attacker intentionally creates a package that contains the same words as the targeted package, but the order of the words will vary. The goal is similar to typosquatting, benefiting from an innocent mistake of a developer. A researcher uploaded typosquatting packages to different repositories, including PyPi, and the package received around 45K downloads over several months \cite{vu2020typosquatting}.

\subsection{Machine Learning}

Learning is considered the hallmark of human intelligence, and researchers have worked on making machines intelligent, giving birth to ML \cite{wang2009brief}. It is the study of methods to allow computers to gain new knowledge and constantly improve themselves \cite{wang2009brief}. The definition can also be the process that allows computers to convert data to knowledge through certain techniques \cite{wang2009brief}. The applications of the ML techniques have been quite popular in recent years, and one of these applications is the area of information security. We will incorporate some ML techniques  into our proposed security remedies to provide a more reliable method of identifying and preventing malicious actors.

\section{Motivation}

Although several studies exist in the field of OS registry and software supply chain security, there has not been a known source to publish a survey-structured paper to construct a centralized information bank on the types of dangers in OS registries. Furthermore, we have not found a project to offer mitigation strategies against these attacks. The feasibility of our solutions may be unclear, as we have not tested any of them. However, we hope they would serve as a first step to develop more systematic countermeasures.

OS registry security is a relatively generic field that tackles certain security issues in OS environments. It quite surprised us that a survey does not yet exist on the security of the public library registries. Therefore, to contribute to the field and give a glimpse of the area's current state, this work has been thought to be useful. We hope to enhance the information security field by offering risk mitigation strategies and serving as a first step to build a novel chapter in its curriculum. We expect to have several corrections done on this draft of a chapter. Still, our ultimate motivation is to foster interest in the field and raise awareness against these issues as we also support the OS community.

\section{General Overview of Vulnerabilities in npm and PyPi}

\begin{figure}[h]
  \centering
  \includegraphics[width=\linewidth]{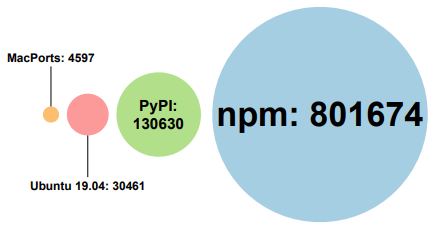}
  \caption{A size comparison in package management ecosystems \cite{vaidya2019security}}
  \label{size-comparision}
\end{figure}

Although several other package managers and OS environments exist, we decided to focus on PyPi and npm as they contain the most libraries in their registries. Similar ecosystems and their size comparisons are illustrated in figure \ref{size-comparision} \cite{vaidya2019security}.

\begin{figure}[h]
  \centering
  \includegraphics[width=\linewidth]{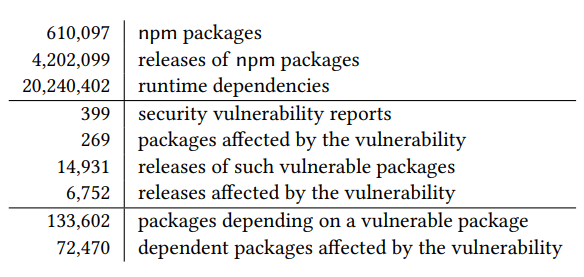}
  \caption{A summary of the npm dataset \cite{decan2018impact}}
  \label{vulnerability-table}
\end{figure}

To understand the scope of the security issue and get an estimation of the number of vulnerable packets, we scanned research papers. We found that one of the projects in our references compiled a summary of 700 security reports gathered from Snyk.io published before 2017-11-09 in figure \ref{vulnerability-table} \cite{decan2018impact}. However, researchers removed certain issues, such as typosquatting, as they do not introduce vulnerabilities in existing packages \cite{decan2018impact}. Among the 610K packages, 133602 packages directly depend on vulnerable packages, and 52\% of these packages, which is 72470 in total, have at least one release that relies on a vulnerable version \cite{decan2018impact}.

The median number of releases affected by a high-severity vulnerable package is 26 in the npm environment \cite{decan2018impact}. The encouragement from the npm and PyPi environment to reuse code seems to accelerate this issue \cite{vaidya2019security}.

\subsection{Direct, Indirect Dependencies, and Heavy Code Reuse}

\begin{figure}[h]
  \centering
  \includegraphics[width=\linewidth]{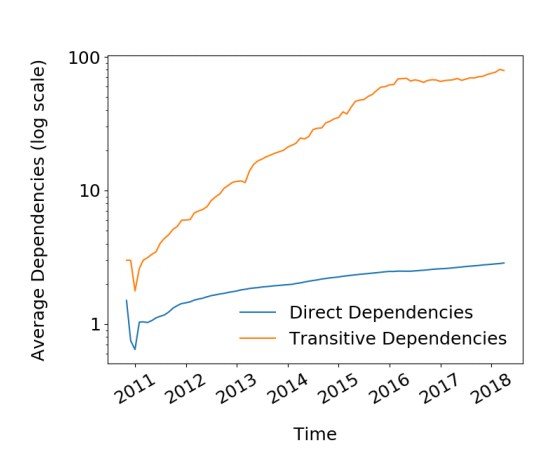}
  \caption{The growth of both types of dependencies over time in logarithmic scale \cite{zimmermann2019small}}
  \label{dependency-growth}
\end{figure}

As stated before, code reuse is common in the npm environment. Projects are either directly or indirectly dependent on each other, meaning that project A that uses code from project B has a direct dependency relationship as A directly depends on B. If B depends on project C, then A has an indirect dependency on C, increasing the attack surface. Indirect dependencies are also referred to as "transitive dependencies" as some of the work in the literature uses that term to build their graphs, such as figure \ref{dependency-growth} that indicates the average number of dependencies projects use over time.

Figure \ref{dependency-growth} implies that a developer implicitly trusts around 80 other packages by using a library \cite{zimmermann2019small}. This trust also means that a hacker's attack surface includes the accounts that maintain the indirectly dependent packages.

\begin{figure}[h]
  \centering
  \includegraphics[width=\linewidth]{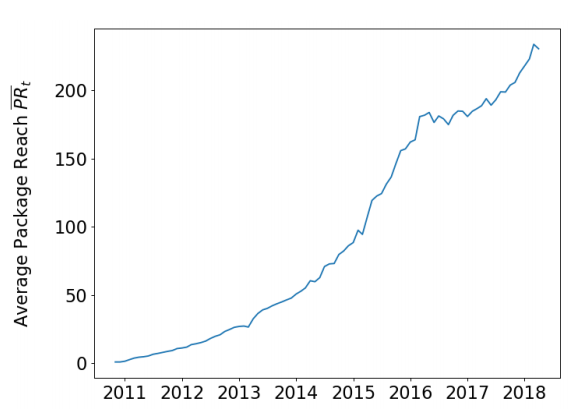}
  \caption{The average package reach overtime in the npm environment \cite{zimmermann2019small}}
  \label{average-package-reach}
\end{figure}

Another term based on transitive dependency is the "Package Reach," which refers to the number of transitive dependencies a specific library has \cite{zimmermann2019small}. It has been stated that some popular packages have a package reach of over 100000, drastically increasing the attack surface, and this trend has been observed to be increasing \cite{zimmermann2019small}. The package reach also increases the size of the dependency graph. If one of the dependent packages is compromised by either a squatting attack or a stolen credentials of a maintainer account, it would be reasonable to assume that a security risk would arise. The average package reach for a single given library seems to be in an increasing trend for the last couple of years, according to figure \ref{average-package-reach}.

\begin{figure}[h]
  \centering
  \includegraphics[width=\linewidth]{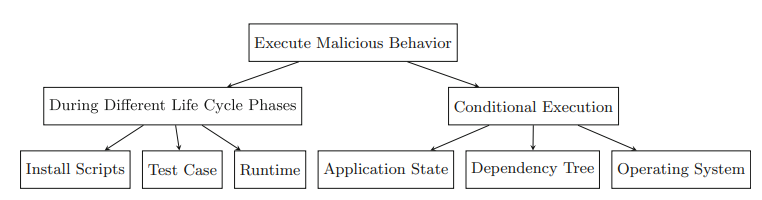}
  \caption{The attack tree that outlines the possible phases a malicious script can be executed \cite{ohm2020backstabber}}.
  \label{attack-tree}
\end{figure}

Once the attacker injects his code into a dependency graph, he can execute it in various phases. The attack tree that outlines wherein an attacker can execute the malicious script is also illustrated in figure \ref{attack-tree}.

Zimmerman's 2019 USENIX paper seems to be a leading study examining the interdependent structure of packages and their trends in the npm environment. However, it has also been determined that most papers in this field seem to be less than five years. This finding implies that security experts recently started to investigate the interdependent packaging structure of OS environments. Software supply-chain attacks, on the other hand, seemed to be a more mature field.

\subsection{Technical Lag}

Packages relying on vulnerable libraries is not the only issue, as Gonzalez-Barahona et al. introduced the concept of technical lag, which is "the increasing lag between upstream development and the deployed system if no corrective actions are taken" \cite{gonzalez2017technical,zerouali2019formal}. It is essentially the phenomenon where a package lags behind its most recent release \cite{decan2018evolution}. Code reuse only seems to be contributing to the issue as most recent packages can still be outdated if they rely on unpatched components \cite{zerouali2019formal}. 

One may reasonably want the package updates to be automated to solve the issue. Package updates may introduce incompatibility issues with the existing project \cite{zerouali2019formal}. Developers should not ignore patches as well due to known vulnerabilities. Patches also introduce new functionalities that a project misses by avoiding incompatibility risks, causing a concept named "technical debt" to arise \cite{gonzalez2017technical}. However, the issue mainly seems to be a trade-off between security and functionality. 

The perspective against technical lag also seems to vary as the literature contains data implying that clients can be safe from technical lag. It has been concluded that one in four dependencies and two in five releases suffer technical lag in the npm environment \cite{decan2018evolution}. After 2015, it has also been observed that the average duration of technical lag in npm was 7 to 9 months, meaning that a specific dependency would be updated 7 to 9 months after a library's new release \cite{decan2018evolution}. 

Furthermore, benign users might not be the ones to discover a vulnerability first. Most vulnerabilities take a long time to be discovered, especially low-severity ones \cite{decan2018impact}. Although project contributors fix most vulnerabilities discovered in the source code in a short amount of time, a non-negligent portion takes longer \cite{decan2018impact}. 

To counter the dangers of the technical lag argument, we found one report stating that 73.3\% of outdated clients were safe from threats as these clients did not use the specific vulnerable code \cite{zapata2018towards}. A vulnerability tends to be in a specific function, and not every client will have to use that particular function. However, the risk of a client using the vulnerable function persists. Developers who realize the functions they are using are not particularly affected by the stated issue take longer to update their dependencies, increasing the technical lag \cite{zapata2018towards}.

In addition, One may reasonably conclude that the older the package is, the less vulnerability it will have. Thus, a possible solution may be to use old packages. However, a study in the literature has found that this claim is wrong as they found that most vulnerabilities are in packages older than 28 months \cite{decan2018impact}. Older packages tend to have more severe vulnerabilities \cite{decan2018impact}. We could not establish a relationship between security and the package's age in this study.

\subsection{Squatting Attacks}

We found that attackers mainly use two methods to spread malicious code in these environments: stealing the credentials of accounts that maintain certain packages to inject their code into the project and tricking users into downloading their packages by methods such as squatting attacks \cite{vu2020typosquatting}. The term squatting attack is an umbrella concept used in this paper to represent typosquatting and combosquatting. We will discuss other types of attack methods against the public registries in this paper, but we focus on these two scenarios in this section. In the first scenario, the “ssh-decorate” package was affected as the attackers took control of the maintainer account and injected code into the package. The injected code would send users' ssh credentials to an attacker-controlled remote server \cite{vu2020typosquatting}. Some popular maintainers can contribute to hundreds of packages, making their accounts critical to a potential compromise \cite{zimmermann2019small}. 

In the latter scenario, attackers can fork an existing package and modify the \textit{setup.py} file to download the malicious dependencies in PyPi \cite{vu2020typosquatting}. In addition, an attacker can even use an existing package name after a popular package has been withdrawn by the maintainer from the registry, introducing further attack surfaces \cite{ohm2020backstabber}. One such attack had happened in the past with the go-bindata package when its owner deleted the unmaintained package {vaidya2019security}. Fortunately, the attack was discovered on the same day \cite{vaidya2019security}. Squatting attacks are a problem in the OS registry environment because the repositories allow users to upload code as users are given equal trust initially \cite{taylor2020spellbound}. 

Squatting attacks are both dangerous for the developer, stakeholders, and the client as a Trojan virus may be embedded that activates when the project is run \cite{taylor2020spellbound}. Certain code runs on users' machines when npm or PyPi installs packages on the local repository. Thus a client does not even have to use the dependency to be attacked. Some packages are even known to open reverse shells with user privileges using the installation scripts of the packages \cite{taylor2020spellbound}. Manual squatting detection proves to be challenging as some of the name changes are quite hard to notice, as illustrated in table  \ref{malicious-package-samples} that shows some past attacks \cite{taylor2020spellbound}. 

\begin{table*}[t]
\centering
\begin{tabular}{ | c | c | c | }
  \hline
 Malicious Package & Legitimate package & Names change \\
 \hline\hline
 virtualnv & virtualenv & Delete ‘e’ \\ 
   \hline
 mumpy & numpy & Substitute ‘n’ by ‘m’ \\ 
   \hline
 django-server & django-server-guardian-api & Delete “-guardian-api” \\  
 \hline
  urlib3 & urllib3 & Delete ‘l’ \\  
 \hline
   python-mysql & MySQL-python & Swap “python” and “mysql” \\  
 \hline
    python-openssl & openssl-python & Swap “openssl” and “python”  \\  
 \hline
\end{tabular}
\
   \caption{Some of the samples of squatting attacks selected from the literature in the PyPi environment \cite{vu2020typosquatting}}
  \label{malicious-package-samples}
\end{table*}

The dangers of a typosquatting attack come from the fact that the client is not affected by the malicious activity only but all the other libraries that depend on it, meaning that a developer does not have to make a typo mistake \cite{taylor2020spellbound}. Another challenge in catching these attacks is that not all typosquatting packages are malicious, as the \textit{loadsh} example did not have any harmful content \cite{taylor2020spellbound}. However, it was a copy of an older version of the original \textit{lodash}, meaning that public unpatched vulnerabilities still may be a security concern, as it has been reported that 63 other packages depended on \textit{loadsh} \cite{taylor2020spellbound}. 

\subsection{Maintainers and Collaborators}

There is a distinction between maintainers and collaborators in OS environments. Maintainers review and approve contributions while contributors propose code changes \cite{ohm2020backstabber}. Both of these accounts are critical for a secure project. An attacker's goal is to inject malicious code into the dependency tree to affect the targeted software. An attacker can reach that goal by either creating a new package via squatting or using the existing packages that already have users \cite{ohm2020backstabber}. If a maintainer account is compromised, an attacker can easily inject his code. Another possible scenario is that an attacker can pose as a benign contributor and make pull requests with a seemingly useful feature, such as the incident of the unmaintained "event-stream" package that will be mentioned in the following sections \cite{ohm2020backstabber,vaidya2019security}. Embedded inside the feature can be malicious content that may be hard to notice, and considering that a human is reviewing the request, human error can allow the request to be approved. Alternatively, a maintainer account with weak credentials can always be compromised to approve malicious pull requests, or attackers can use social engineering strategies to perform the injection.

\begin{figure}[h]
  \centering
  \includegraphics[width=\linewidth]{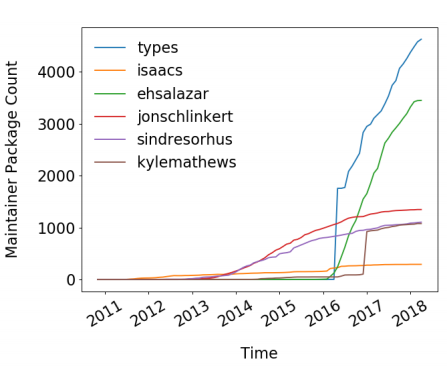}
  \caption{Package count per popular username in npm \cite{zimmermann2019small}}
  \label{package-count-per-popular-maintainer}
\end{figure}

The trend presented in the transitive dependency structure is not the only alarming graph. The rise of npm would be from new developers joining the system mainly; however, the package per maintainer has also increased as current members publish new projects \cite{zimmermann2019small}. Each account is becoming responsible for more packages, making it more critical to secure accounts. The average package count per maintainer account seems to be 2.5 in 2012, ascending to 3.5 in 2013 and almost 4.5 in 2018 \cite{zimmermann2019small}. This increase happened for both average and especially for influential maintainers in npm, as the case for some selected popular maintainer usernames can be observed in figure \ref{package-count-per-popular-maintainer} \cite{zimmermann2019small}.

\subsection{Trivial Packages or Micropackages}

Trivial packages are small libraries, which a study in the literature found to be less than 35 lines of code, according to 79\% of their participants who attempted to classify libraries as trivial \cite{abdalkareem2020impact}. Developers have considered trivial packages good until recent trends pushed code reuse to an extreme \cite{abdalkareem2020impact}. The breakdown of popular web services, such as Facebook, Airbnb, and Netflix in 2016 from a Node.js 11-line trivial package named left-pad, only made the concept be questioned further \cite{abdalkareem2017developers,chen2021helping}. 

The incident has been referred to as the case that "almost broke the Internet," which led to many discussions over code reuse sparked by David Haney's blog post "Have We Forgotten How to Program?" \cite{abdalkareem2017developers}. Node.js used to allow developers to unpublish projects that lead to the left-pad incident emerge \cite{abdalkareem2017developers}. Although the origin of the incident is not related to a vulnerability in the package itself, it has raised awareness about trivial packages \cite{abdalkareem2017developers}. Numerous developers agreed with Haney's opinion that developers should implement trivial tasks themselves rather than adding an extra dependency to the project \cite{abdalkareem2017developers}. Since then,  people have been working to investigate this issue.

A study has researched the developers' perspectives of trivial packages or micropackages, and the interviewed developers stated that their definition of micropackage is the same across PyPi and npm \cite{abdalkareem2020impact}. It has been found that 16\% of packages in the npm and 10.5\% of packages in PyPi are trivial packages \cite{abdalkareem2020impact}. The same study also questioned developers to understand the reasons for using such packages. It has been found that developers use these packages due to their belief that the packages are well tested and maintained \cite{abdalkareem2020impact}. Developers stated that they use these packages to increase their productivity \cite{abdalkareem2017developers}. The surveyed software developers also stated that they use trivial packages as they do not want to be concerned with extra indirect dependencies. However, it has also been found that trivial packages do use dependencies in some cases \cite{abdalkareem2020impact}. A study has found that 43.7\% of trivial packages have at least one dependency, and 11.5\% have more than 20 dependencies \cite{abdalkareem2017developers}. The surveyed developers also thought that trivial packages create a dependency mess, which is hard to update and maintain \cite{abdalkareem2017developers}. The perspective against trivial packages seems to be controversial, as it has been found that 23.9\% of the JavaScript developers consider them bad. In comparison, 57.9\% of the developers do not consider them to be a bad practice \cite{abdalkareem2017developers}. 70.3\% of the Python developers consider trivial packages as bad practice \cite{abdalkareem2020impact}.

Zimmermann's paper coins this issue as "micropackages" instead of using the term "trivial package." Still, the concepts are the same: packages with fewer lines of source code than a threshold, although this threshold seems ambiguous across the literature \cite{zimmermann2019small}. The specific study explicitly stated that micropackages are insecure as it increases the attack surface and the number of dependencies a project has \cite{zimmermann2019small}.

\subsection{PyPi Overview}

\begin{figure}[h]
  \centering
  \includegraphics[width=\linewidth]{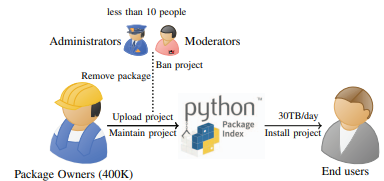}
  \caption{An overview of the roles in the PyPi ecosystem \cite{vu2020typosquatting}}
  \label{pypi-ecosystem}
\end{figure}

%% TODO: I assume this is only vulnerabilities for PyPI but not npm? Also should we name it as PyPi-Specific Vulnerabilities?
PyPi has limited automated review tools for the uploading process as the npm environment does, making it vulnerable to different kinds of attacks, such as squatting \cite{vu2020typosquatting}. Furthermore, the moderator and administrator team, who has permission to remove packages from the registry, seems to be less than ten people, limiting the maximum number of code reviews they can conduct \cite{vu2020typosquatting}. Considering the 400K package owner, each administrator seems to be responsible for 40K people, assuming every administrator performs code review for malicious content, thus providing a lower bound for the number of package owners per moderator ratio. PyPi allows end-users to report malicious packages. Nevertheless, considering each moderator being responsible for at least 40K developers, it would be only reasonable to question the efficiency of the code reviews. When users download packages using the \textit{pip}, there is no available system that reviews the code to determine its safety aside from a user's antivirus. So, we can outline the process of publishing packages with figure \ref{pypi-ecosystem} that illustrates a high-level view of the schematics of the PyPi ecosystem.

Spreading malicious code in the wild PyPi is fairly similar to the process we illustrated in the npm environment. Mainly, an attacker can either steal the credentials of an existing account to exploit the current reputation of the project or create a new package by forking the targeted package and modifying the content, or simply creating a brand new package \cite{vu2020typosquatting}. The latter method can still trick users into downloading the attacker's library by a squatting method.

\subsection{Noteworthy Incidents}

Although we described the attack types, we have not given detailed concrete examples until this point. Thus, it would be fruitful for the reader to illustrate these concepts with real-life scenarios.

%% TODO: could we provide a better ordering of these incidents. Or we could also add a summary line before incidents, especially for the first incidents. Is there a way to relate it with vulnerabilities we have mentioned in previous sections?
In July 2018, an attacker stole the credentials of a developer account of a package. The attacker published a malicious version of the \textit{eslint-npm} package that would, during installation, send users' npmjs.org credentials from the .npmrc file to an attacker-controlled server\cite{vaidya2019security}. Although the attack was detected the same day, it has been estimated that 4,500 accounts may have been compromised \cite{vaidya2019security}. This incident clearly shows the importance of the period until an attacker is detected. 

A known typosquatting attack happened in July 2017 when a user named "HackTask" uploaded 40 malicious packages to the npm registry with similar names to popular packages. The packages had payloads in their installation script that would ex-filtrate local environment variables, storing sensitive authentication tokens, to the attacker's server \cite{vaidya2019security}. The attack was discovered after 12 days, and it has been estimated that 50 users downloaded the attacker's packages \cite{vaidya2019security}.

Another typosquatting attack happened in the PyPi public registry, where a malicious package, named \textit{jeIlyfish}, hoped to imitate the popular package \textit{jellyfish} by changing the first L of jellyfish to an I, and it stole ssh and GPG keys \cite{vu2020towards}. The malicious package existed in the repository for a year until it was discovered, thus implying that squatting attacks do not necessarily get discovered in a short period \cite{vu2020towards}. One study, which also expressed that vulnerability detection and prevention are insufficient in PyPi, found that the median time a vulnerability is hidden in the PyPi environment is three years regardless of their severity\cite{alfadel2021empirical}. Once the vulnerability is discovered, on median, all packages take four months to have their vulnerability patched, giving plenty of time for an attacker to exploit a zero-day vulnerability \cite{alfadel2021empirical}.

A social engineering technique was also used to attack the npm ``copay" package, and the incident dates back to July 2018, when the attacker emailed the maintainer to offer help with the popular but unmaintained "event-stream" package that copay has an indirect dependency on \cite{vaidya2019security}. Once the attacker gained access to contribute, the attacker injected a malicious dependency into an external package, ``flatmap-stream," of event-stream that would ex-filtrate sensitive wallet information of the copay users to the attacker's server \cite{vaidya2019security}. 

All in all, there have been several examples of attacks in npm and PyPi. The interdependent structure of both environments raises software supply-chain security as a concern. It is not only npm and PyPi that gets targeted but other popular package manager software that we have not covered in this survey. One such example is the popular \textit{rest-client package} from the RubyGems registry. attackers used it to insert a Remote-Code-Execution backdoor on web servers \cite{duan2020measuring}. A study in the literature stated that the number of these attacks increases \cite{vu2020towards,alfadel2021empirical}. Therefore, we expect to see additional incidents in the field if admins do not implement countermeasures. 

\section{Discussion}

This research focused on presenting the costs of using OS registries to benefit from the readily available libraries other developers have created. These costs mainly originate from security concerns that may damage institutions' IT infrastructure. Evidence has been gathered from the literature to study the main threats in the npm environment comprehensively. No further experiments or studies have been done in this work, as we solely focused on compiling recent work in OS registry security. We determined that the OS environment provides new attack surfaces to inject code into unsuspecting victims’ machines. An attacker injecting a script into a victim’s computer can have devastating effects, considering the Turing-completeness of these languages. 

\subsection{Suggested Countermeasures}

We have developed countermeasures based on the ideas we gained from our literature search. We specifically attempted to incorporate ML techniques into our approach. We took this step as the offered solutions will only need minor modifications to features to cover new undiscovered future attacks. This advantage of ML will make the presented techniques more adaptable. Furthermore, although signature-based malware detection is the most extended technique in commercial antivirus, they tend to fail as newer types of malware emerge \cite{santos2013opem,scott2017signature}. Artificial intelligence provides a competitive advantage over next-generation malware \cite{scott2017signature}. 

Therefore, when designing our countermeasures for public registries, we incorporated ML techniques. We have not tested the solutions yet, so the feasibility of using them may be in question. Another option can be contacting the admin teams of the public registries to gain their inputs. However, we hope that this work would act as a first step to spark interest in this field and capture the attention of the admin teams of OS registries.

\subsubsection{Malicious Package Detection}

Some sources state that current tools for identifying malicious payloads are resource-demanding. However, there are lightweight tools in the literature to mitigate the risk of malware residing in the registries. These tools can be automated scanning programs to analyze the dependency tree for squatting packages \cite{vu2020towards}. The automation of reviewing published code may prove to be challenging. Nevertheless, allowing any developer to publish code into the public registry, where millions of developers have access, gives much power to any stranger. 

Since the source code for every project on an OS registry is public, any user can review the code for each library. This fact is advantageous as the dependency tree and the source code of the project is visible. One antivirus designed for APIs can review packages on the registry to eliminate the time needed for the public to uncover malicious software. A member of the admin team or a volunteer can download batches of packages from the public registry to his local repository and use the antivirus to ensure the safety of each library. However, we have not found an antivirus software specifically designed to scan APIs on the internet. Antiviruses may use features that work on software for end-users. The same features may not prove efficient for libraries designed for implementing programs. 

For this purpose, we will explore antivirus software that may be useful specifically for APIs. We will focus on analyzing the binary and the behavior of the library instead of its source code. We decided to cover a broader range of use cases, as we believe analyzing the source code is less challenging than an executable binary. Thus, our project can also cover library binaries for malware detection.

There has been work in the field to efficiently automate the detection of malicious code injection in the distributed artifacts of packages, and the admins may attempt to implement some of these novel tools \cite{vu2020towards}. One such tool, named \textit{Buildwatch}, analyzes the third-party dependencies by using the simple assumption that malicious packages introduce more artifacts during installation than benign libraries \cite{ohm2020towards}. This hypothesis has been formulated and tested in the \textit{Buildwatch} study \cite{ohm2020towards}. Admins can also modify \textit{Buildwatch} to detect squatting packages in the dependency tree with the techniques we will discuss in the upcoming sections.

One approach in detecting suspicious packages comes from a study that uses anomaly detection to identify suspicious software \cite{garrett2019detecting}. Their approach first extracts several features from the version when a package gets updated. It then performs k-means clustering to detect outliers for further review. The researchers kept the number of features low to ensure the clustering algorithm takes limited computation power, thus making it lightweight \cite{garrett2019detecting}. Some extracted features are the ability to send/receive HTTP requests, create/read/write to file systems, or open/listen/write to sockets \cite{garrett2019detecting}. This tool specifically uses anomaly detection, a branch of ML, to tackle the issue of detecting suspicious updates. We believe this tool is very suitable for analyzing third-party libraries.

As image classification becomes more accurate due to the development of robust classifiers such as convolutional neural networks (i.e., CNN), researchers in the field proposed visualization-based techniques to detect malicious packages. This technique would input the executable of the library to visualize its binary for processing. An advantage of binary visualization is that it does not need to disassemble the package to perform static analysis if the source code is not available. The program can directly examine the package binary. 

Nataraj et al. ~\cite{nataraj2011malware} proposed a method to convert the malware binary into a gray-scale image by reading the binary as vectors of 8 bits and organizing them into a 2D array. They extracted texture features from the image representation of the malware and performed K-nearest neighbor (i.e., KNN) with Euclidean distance to classify the malware. 

Ni et al.~\cite{ni2018malware} converted malware opcode sequences into gray-scale malware fingerprinting images with SimHash encoding and bilinear interpolation and used CNN to train the malware classifier. Their evaluation results revealed that their classifier outperformed support vector machine (SVM) and KNN on the features from 3-grams of the opcode, and another classifier using KNN on GIST features. The classifier reaches around 98\% accuracy and only takes 1.41s to recognize a new sample. 

Besides the visualization of the binaries of libraries, which is similar to static analysis, there are other ML-based techniques. The ML-based malware detection field seems well-established as several survey papers exist, such as studies from Gandotra et al. and Ucci et al. \cite{gandotra2014malware,ucci2019survey}. They compiled various tools from the literature, and Gandotra et al. listed their limitations \cite{gandotra2014malware}.  These tools are based on static analysis, in which the code is not executed, and dynamic analysis, which focuses on the run-time behavior \cite{gandotra2014malware}. One such tool even uses data mining techniques to detect malicious executables that rely on specific features from the binary. \cite{schultz2000data}. These features are the Portable Executable, strings, and byte sequences \cite{schultz2000data}. The Naive Bayes model takes in these features and reaches an accuracy rate of 97.11\% in the study \cite{schultz2000data,gandotra2014malware}. These surveys contain several examples of malware tools that can be adapted for APIs.

In 2014, Gandotra et al. stated that many researchers lean towards dynamic analysis instead of static \cite{gandotra2014malware}. The survey gave examples of dynamic analysis techniques for malware detection \cite{gandotra2014malware}. Some examples include the work of Rieck et al. that monitors the behavior of malware on a sandbox environment \cite{gandotra2014malware,rieck2011automatic}. The observed behavior of the program is embedded into a vector space \cite{rieck2011automatic}. They use clustering to identify similar malware behaviors to classify each executable \cite{rieck2011automatic}.

To conclude the antivirus examples summary for APIs, we believe that the volunteer developers recruited via an advertisement banner on the official npmjs.com website can implement the mentioned tools and ML techniques in exchange of the trust score incentive in the following sections. Afterward, the admin team can design a protocol where a volunteer user can install batches of libraries into his local repository to allow the mentioned programs to review their safety. Developers can even automate this protocol from a web server to prevent the loss of precious human labor on repeated tasks. 

If our methods are not feasible for public registries, novel tools still exist in the literature accessed from the Google Scholar platform. We could not find existing antivirus explicitly designed for libraries, but already-tested software would be a better option if admins can access it. Admins can decide to move forward with a novel tool originated from a paper, and reading academic papers can indeed be a daunting task as it is for us. However, most papers have correspondent authors that can clarify ambiguous implementation details. Some papers even have their implementations on public repositories, which will be very helpful to the team of implementers.

\subsubsection{Reasonable Constraints, Decentralization, and Trust Score}

%% TODO: is there ways to include ML techniques here? For instance, detect malicious maintainer, collaborators, or commits using ML. Or maybe use ML to handle voting and filter untrustful scores from untrustful users?
The founding committee of an OS registry might not operate to pursue profit, and they may be short on budget to hire additional developers to implement these tools or reviewers to find malicious packages manually. However, to solve this problem, the admin teams can decentralize the manual code review process to people willing to contribute to the platform's security. Contributors can opt-in to review code and cast votes. After a new library is reviewed and voted to be published by a certain number of developers, it may be available to install on the public registry. 

To encourage members of the OS registries to become reviewers, admin teams of the registries can give specific incentives for the time a reviewer invests. These perks, such as early access or discounts to products, can be supplied by companies that sponsor the npm or PyPi environment. As reviewers cast their votes to pass or reject a commit, they can earn points to unlock some advertised perks. The point system can be gamified to build public rankings and hierarchic levels. Each level would give contributors different perks, such as a free course on a partnered educational platform. 

To implement the final stage of the voting-based solution, one has to take measures against the possibility of reviewers abusing the scoring system by casting votes without properly reviewing a commit. We believe there are already known solutions for this issue that does not need to be mentioned in this work. Some of these control measures can include comparing the time it takes for each reviewer to cast a vote to ensure each reviewer invests a proper amount of time. Simply observing the number of times a reviewer is an outlier based on the majority of the votes cast for a certain decision can also be observed. These solutions are merely to preserve the quality of the reviewer committees. 

The scoring system can be implemented to give more privileges to users gradually besides the mentioned perks. Users can earn trust points as they actively review the contributions to projects, and users may start to publish their packages after their score passes a threshold. Thus, registries can give power to partially trusted users rather than anyone with an internet connection. We will be referencing this trust score in some of the following sections.

\subsubsection{Securing Critical Accounts}

We believe the admin teams should implement a mandatory feature in the registries. A user who maintains a popular library or a popular library that uses the user's package should automatically have multi-factor authentication (MFA) to mitigate the risk of compromised accounts. MFA should be mandatory in highly trusted accounts that pass a preset score threshold as well. Any abnormal behavior, such as login from different machines, should be recorded and processed accordingly to prevent more sophisticated attacks. This process can simply be a secret question or a one-time code.

Another possibility is integrating code authorship identification systems into OS registries, where the program would get triggered whenever a code is published from a known author. Known hackers can be profiled from their public source code to create a database of dangerous code styles that may indicate compromised accounts. This practice is a research field in itself, and it would require high investment to build such an infrastructure. However, there are already systems built to tackle this issue to an extent. One such system is demonstrated in a study that demonstrates a Deep Learning-based Code Authorship Identification System (DL-CAIS), which is language oblivious and resilient against code obfuscation \cite{abuhamad2018large}. DL-CAIS uses GitHub repositories as its dataset and reaches an accuracy of 96\% \cite{abuhamad2018large}. It first preprocesses the input code to build representation models and feeds those models into the deep learning architecture to find more distinctive features \cite{abuhamad2018large}. The architecture to be trained consists of several RNN layers and fully connected layers \cite{abuhamad2018large}. After the training, researchers use the resulting representation models to construct a random forest classifier \cite{abuhamad2018large}.

Though limiting ourselves to only known hackers may be an insufficient approach. However, npm and PyPi have the source code of each project, proving to be very advantageous to profile developers. npm or PyPi can use code authorship detection tools to profile each of its developers. Each constructed profile can be associated with a collaborator or maintainer account. If the uploaded code profile in one commit does not match the code profile of the account, the admin can place a temporary hold on the account for further verification. One may still be concerned about the privacy aspect of this approach, but the feature can be optional for non-critical accounts. An automated script can reward users with an extra trust score for opting into the program to provide incentives for this practice. However, we believe critical accounts should be required to participate, as the cost of an account compromise would be significantly higher in their case. 

\subsubsection{A Proposed Solution for Technical Lag}

One feature in the npm environment we would like to see removed is the ability to constrain the specific version of a dependency by default in the package.json file \cite{zimmermann2019small}. We believe that there should be a special keyword in package.json, such as \textit{xxx}, for the npm to check the most updated version in its registry and install it automatically to mitigate the issue of technical lag. The keyword \textit{npm outdated} may be unknown to some developers, or the additional workload it would add can be a deterrent factor. However, the keyword that indicates the most recent version should be a default or auto-filled by the package manager to avoid the user interfering with versions. However, this also presents the danger of an update breaking existing systems. The user should always have the option to fix the dependency to a specific version as some releases will provide long-term support or better stability. Thus, the user can have the freedom to decide if the security or functionality of the project is a priority.

Frameworks have also been developed to measure how outdated a system is, balancing security and functionality \cite{gonzalez2017technical,zerouali2019formal}. Calculating a project's recentness score may prove difficult, as each project will have its priorities, such as performance or stability \cite{gonzalez2017technical}. Gonzalez et al. name this priority as "gold standard" and defines the gold standard to build a lag function that can output the recentness score \cite{gonzalez2017technical}. For instance, if the user chooses his gold standard as security, the lag function can focus on the number of new updates related to security \cite{gonzalez2017technical}. If the gold standard is functionality, the lag function can take the number of new features into account \cite{gonzalez2017technical}. Gonzales et al. even propose other possible lag functions, such as one that simply measures the differences in lines of source code between the most recent and currently deployed dependency or the number of commits of difference between them \cite{gonzalez2017technical}. 

To partially implement the framework in npm or PyPi registries, they can mandate reputable libraries to list and classify each update done with the commit. Users can specify their "gold standard" with a specific command after the creation of their projects. npm can use this gold standard to compare the score of the current dependencies with their most recent versions. Now, the user will know his dependencies' recentness in terms of his priorities. Thus, the user can make more informed decisions on whether he should update.

Registries do not have to undergo significant changes to calculate the current projects' scores, as npm can simply take their score as 0. If all future versions require the contributor to classify new features or bug fixes manually, we believe it will not be challenging to implement the new system. Currently, we do not expect to calculate the significance of each particular update, but the user can be asked to assign a score of significance to each bug fix or new feature. We expect several users to avoid spending time with the classification of new features; however, npm can provide incentives to encourage users to participate in the program, such as with extra perks or trust scores.

\subsubsection{Automatic Eviction of Squatters}

%% TODO: Talk about ML based defense techniques
The brute force solution of this problem can be a function that compares the name similarity of a newly published package to an already existing library to determine suspicious packages. Any new package upload can trigger the npm registry to invoke a script that compares the uploaded library name to every other package name in the registry. Suppose the similarity result passes a certain threshold. In that case, the registry can automatically reject the uploaded package, informing the publisher that the package name is too similar to an existing one, and he should choose a new name. However, this approach may prove to be insufficient as the script may not be able to capture every possible scenario.

A simple countermeasure in the literature is the tool TypoGard. It implements a novel detection technique, which uses both the lexical similarity of the package names and the package popularity \cite{taylor2020defending}. It flags certain cases where a user would download an unknown package with a similar name to a popular one before the library can be installed \cite{taylor2020defending}. The tool even caught previously unknown attack attempts, including a package named \textit{loadsh} that imitates the popular npm package \textit{lodash} \cite{taylor2020spellbound}. An advantage of TypoGard is that it can simply run on the client-side and requires no cooperation from repositories \cite{taylor2020defending}. Researchers evaluated TypoGard on npm, PyPi, and RubyGems, and it has achieved an accuracy of 99.4\% with flagging cases \cite{taylor2020defending}. The researchers also measured the overhead of TypoGard and determined it is only 2.5\% of the install time \cite{taylor2020defending}. We found that TypoGard is also known as SpellBound in the literature \cite{taylor2020spellbound}.

Squatting also exists in DNS servers, and researchers proposed ML-based solutions to identify suspected squatters in the past \cite{moubayed2018dns}. One study proposes approaches based on supervised learning. It specifically uses an ensemble learning classifier that uses various algorithms, such as SVM, that reach an accuracy rate of 88.4\% \cite{moubayed2018dns}. The study also uses unsupervised learning to identify suspicious domain names by implementing K-means clustering based on specific futures extracted from domain names \cite{moubayed2018dns}. Here, the key takeaway seems to be the features chosen to reach this accuracy, such as the number of unique letters or unique numbers in a domain name \cite{moubayed2018dns}. For the OS registries case, developers tasked with implementing countermeasures can examine past squatting attacks to create potential features to identify squatters. The features are mostly derived initially by making assumptions of what commonalities of past squatter packages have. These features can even be created by guessing, then feeding them to the ML algorithm to observe which produces the highest accuracy rates.

\subsection{Future Direction}

First and foremost, all of our suggested countermeasures are theoretical as none of them has been tested in a real-world scenario. We can do additional work by reaching out to the OS registries' admin teams to conduct interviews on the feasibility of our solutions. The admin teams can also have the chance to voice their opinions and common issues they encounter. Based on the issues OS registries face, we may modify some of our solutions or search for additional tools in the literature to meet the admin team's requirements.

%% TODO: maybe we could also discuss the advantages and disadvantages of automatic countermeasures (i.e., ML) and manual/semi-automatic countermeasures (e.g., manual score, opt-in critical account...). We could focus on why ML is better or might improve manual countermeasures. 

%% TODO: ML adversarial example in Binary Visualization
To apply ML in the real world for malicious package detection, we need to ensure that the ML model is robust enough against adversarial examples. Adversarial examples are carefully crafted samples to fool the ML model. Previous work has already illustrated how a clever attacker can augment the correctly classified malware with malicious components to cause the ML model to misclassify it~\cite{khormali2019copycat,kolosnjaji2018adversarial,demetrio2020adversarial,sharif2019optimization}. These attacks mostly inject malicious components to the non-reachable parts of the malware, such as padding bytes to the end of the malware or injecting malicious payload to the malware's DOS header. In this way, they can ensure that the original malware purpose is still maintained. However, these attacks can be filtered easily by embedding a code cleaning process into the malicious package detector to filter out non-reachable codes. A more powerful attack can also inject malicious components into the code section~\cite{sharif2019optimization}. These attacks are more difficult to detect, and further study is required to develop defenses against them. 

%% TODO: Apply ML to code review for reported packages, and also finding similarly malicious packages within the registry based on the knowledge of some known malicious packages (e.g., clustering).
A useful approach to ease the burden of manual code reviewing for uploaded packages is to measure their similarity to the previously reviewed packages. Suppose the uploaded package is similar to a known malicious package. In that case, it might raise a warning immediately, which can help the reviewers decide whether to reject it into the registry. ML approaches are well suited for detecting code similarity. Previous work provides several ML approaches to detect code similarity in binary-level~\cite{shalev2018binary,wang2017memory,marastoni2018deep} or decompiled code level~\cite{ullah2021software}. Existing use cases of the proposed methods focus more on detecting software plagiarism and IP theft. Other use cases such as detecting malicious packages are still worth exploring. 

Another possible future direction is a project focused on the cost-benefit analysis of using OS package registries. Different organizations with various sizes, fields, or interests may have outweighing benefits or costs using the OS ecosystem, and it is worthwhile to examine the analysis. The aspects of OS security do not seem to be either white or black, and sometimes the risks may justify the benefits for a particular use case.

Future work can also extend the survey to other OS tools aside from public registries. Several companies create internal versions of OS projects to avoid potential hazards. However, these internal versions are closed source, and the OS community cannot benefit from additional features implemented within the firm. Informing the academic society about the security hazards of other OS tools and providing risk mitigation strategies may attract businesses to adopt and actively maintain these projects, thus benefiting the OS community. 

\section{Conclusion}

It seems that research cannot completely remove the risks of npm and PyPi as OS projects mainly rely on volunteer humans and public access. It would be justifiable to assume that these factors constantly create new attack surfaces. The OS registries are open to the public and maintained by humans; however, risk mitigation strategies can be implemented, such as requiring the contributors to use MFA or implementing a trust-score system. It is essentially a trade-off between convenience and security. Nevertheless, the possible opportunity cost of the lack of security must be deliberated carefully before reaching a decision.

One must consider the needs of every individual and establishment when developing software. Certain companies cannot risk a possibility of a data breach as their reputation loss in monetary terms would be much larger than maintaining an army of software developers to convert OS tools to closed source. The lack of trust in the registries would deprive the OS community of the potential knowledge the employees of these firms can bring into the field. A concrete example of this issue is that the Department of Defense is concerned about certain dependencies of their projects being out of their control. This concern would justifiably cause them to refrain from using OS platforms \cite{ellison2010evaluating}. With countermeasures born from OS security research, admin teams can minimize the risk. The risk can be minimal to the point where some high-security clearance companies may use and even contribute to certain OS packages. Thus, they will strengthen the OS society even further and make software accessible to anyone.

\bibliographystyle{splncs04}
\bibliography{biblio}
\end{document}